\documentclass[oldversion]{aa}
\usepackage{graphicx}
\usepackage{txfonts}
\usepackage{amsfonts}
\usepackage{subfigure}
\usepackage{amssymb,amsfonts}
\usepackage{natbib}
\bibpunct{(}{)}{;}{a}{}{,}

\begin{document}

\title{
High contrast optical imaging of 
companions: \\ the case of the brown dwarf binary HD130948\,BC}

\author{L. Labadie\inst{1,2}, R. Rebolo\inst{1,6}, I. Vill\'o\inst{3}, J. A. P\'erez-Prieto\inst{1}, A. P\'erez-Garrido\inst{3}, S. R. Hildebrandt\inst{4} \\B. Femen\'ia\inst{1,2}, A. D\'iaz-Sanchez\inst{3}, V.~J.~S. B\'ejar\inst{1,2}, A. Oscoz\inst{1}, R. L\'opez\inst{1}, J. Piqueras\inst{5}, L.~F. Rodr\'iguez\inst{1} 
}

\offprints{{}L. Labadie\\
           \email{labadie@iac.es}}

\institute{
Instituto de Astrofisica de Canarias, C/ Via Lactea s/n, La Laguna, Tenerife E-38200, Spain
\and
Departamento de Astrofisica, Universidad de La Laguna, 38205 La Laguna, Tenerife, Islas Canarias, Spain
\and
Universidad Politecnica de Cartagena, Campus Muralla del Mar, Cartagena, Murcia E-30202, Spain
\and
Laboratoire de Physique Subatomique et de Cosmologie, 53 Avenue des Martyrs, 38026 Grenoble, France
\and
Max-Planck-Institut für Sonnensystemforschung, Max-Planck-Str. 2, 37191 Katlenburg-Lindau, Germany
\and
Consejo Superior de Investigaciones Cientificas, Spain. 
}

\date{Received; Accepted}

\abstract{
{\it Context.} High contrast imaging at optical wavelengths is limited by the modest correction of conventional near-IR optimized AO systems. 
We take advantage of new fast and low-readout-noise detectors to explore the potential of fast imaging coupled to post-processing techniques to detect faint companions of stars at small angular separations. \\
{\it Aims.} We have focused on $I$-band direct imaging of the previously detected brown dwarf binary \object{HD\,130948\,BC}, attempting to spatially resolve the L2+L2 system considered as a benchmark for the determination of substellar objects dynamical masses.\\
{\it Methods.} We used the Lucky-Imaging instrument FastCam at the 2.5-m Nordic Telescope to obtain quasi diffraction-limited images of HD\,130948 with $\sim$0.1" resolution. In order to improve the detectability of the faint binary in the vicinity of a bright ($I$=5.19$\pm$0.03) solar-type star, we implemented a post-processing technique based on wavelet transform filtering of the image which allows us to strongly enhance the presence of point-like sources in regions where the primary halo generally dominates. \\
{\it Results.} We detect for the first time the binary brown dwarf HD\,130948\,BC in the optical band $I$ with a SNR$\sim$9 at 2.561$^{\prime\prime}$$\pm$0.007$^{\prime\prime}$ (46.5\,AU) from HD\,130948\,A and confirm in two independent dataset (May 29 and July 25 2008) that the object is real, as opposed to time-varying residual speckles. We do not resolve the binary, which can be explained by astrometric results posterior to our observations that predict a separation below the telescope resolution. We reach at this distance a contrast of $\Delta I$=11.30$\pm$0.11, and estimate a combined magnitude for this binary to $I$=16.49$\pm$0.11 and a $I$-$J$ colour 3.29$\pm$0.13. At 1$^{\prime\prime}$, we reach a detectability 
10.5 mag fainter than the primary after image post-processing.
\\
{\it Conclusions}: We obtain on-sky validation of a technique based on speckle imaging and wavelet-transform post-processing, which improves the {\it high contrast} capabilities of speckle imaging. The $I$-$J$ colour measured for the BD companion is slightly bluer than, but still consistent with what typically found 
for L2 dwarfs ($\sim$3.4--3.6).
}
\authorrunning{L. Labadie et al.}
\titlerunning{High contrast optical imaging of the brown dwarf binary HD130948\,BC}

\maketitle

\section{Introduction}\label{intro}

A direct determination of dynamical masses of very low mass (VLM) objects is essential to calibrate the mass-luminosity relationship. This is particularly relevant for 
understanding brown dwarfs (BDs) evolution. 
Dynamical masses can be determined by observing close multiple BD systems \citep{Zapatero2004,Bouy2004,Stassun2006,Dupuy2009}. 
BD close binaries with orbital periods  $\lesssim$10\,yr represent a precious sample for a model-independent mass determination within a realistic time baseline. 
Observationally, this requires to spatially resolve the binary, which also permits us to obtain a direct measurement of the flux of each component.  
Since BD systems are also detected as close companions to bright main-sequence stars, another difficulty resides in the strong contrast needed to detect them (cf. the case HR\,7672\,B in \cite{Liu2002}), on top of the detectability issue due to their intrinsic low luminosity. 
So far, the sample of such companion BD binaries is limited to a few number \citep{Burgasser2005}, mostly characterized in the near-IR with the help of 8-10\,m class telescopes. \\
Optical data are necessary for a full characterization of the spectral energy distribution, key to the determination of effective temperatures and bolometric luminosity. In the visible domain, close binaries can be spatially resolved using speckle imaging \citep{Law2006b}, a technique that delivers diffraction-limited optical counterpart to AO-assisted infrared images. The question of {\it high contrast} in speckle imaging has been investigated in the past by \cite{Boccaletti2001} using the ``dark speckles'' method as an additional stage of cleaning to improve the detectability of faint companions. Coupled to the adaptive optics system ADONIS and a Lyot stellar coronograph, these authors obtained K-band contrasts of 1.5--4.5$\times$10$^{-3}$ ($\Delta m_K$$\sim$6--7) at 0.5--0.9$^{\prime\prime}$. \\ 
In this paper, we have focused on the brown dwarf binary HD\,130948\,BC, originally reported by \cite{Potter2002}. As part of a restricted sample of BD binaries companion to a solar-type star, HD\,130948\,BC is a unique benchmark for the study of the mass, luminosity and age of L-type substellar objects. 
Lying at $\sim$\,45\,AU from HD\,130948\,A, the determination of the close BD binary orbit based on infrared AO images permitted \cite{Dupuy2009} to derive a total mass estimated to 0.109$\pm$0.03 $M_{\odot}$. We intent to use speckle imaging techniques to conduct a first optical ground-based high resolution study of this object in order to extend its physical characterization to shorter wavelengths, and at the same time using it as an observational testbench for diffraction-limited imaging from the ground at visible wavelengths. 
Here, we report first $I$-band high contrast speckle imaging of the \object{HD\,130948} system. The detection of the BD companion is improved using image post-processing. This emphasizes how small-medium size telescopes can still be exploited for high angular and high contrast imaging. 
In $\S$ 2 are presented our observations and the data reduction procedure. In $\S$ 3 we present our imaging, astrometric and photometric results on HD\,130948\,BC, briefly discussing our derived $I-J$ colour. 

\section{Observations and Data Reduction}

\subsection{Imaging}

HD\,130948 was observed on the nights of 2008 May 29 and 2008 July 25 using the FastCam instrument \citep{Alex2008} installed at the 2.5-m Nordic Telescope (NOT) at the {\it Roque de los Muchachos} Observatory, Spain. Observing conditions during these nights were good with an average seeing of 0.5$^{\prime\prime}$ in the $I$-band and clear weather. 
In brief, FastCam is an optical imager based on a conventional low-noise CCD camera from Andor Technology that allows us to record speckle-featuring unsaturated images at a rate of several tens of frames per second. Each frame captures a different pattern introduced by the atmospheric turbulence, where each speckle represents a diffraction-limited image of the source of interest. \\ 
FastCam was installed on the Cassegrain focus of the NOT. The pixel scale, determined from astrometric measurements in the M15 globular cluster, is estimated at 31.17$\pm$0.03 mas/pixel. 
The field-of-view accessible with the 512$\times$512 CCD array is 16$^{\prime\prime}$$\times$16$^{\prime\prime}$. 
The raw data acquired by FastCam are composed of cubes of 1000 images each, with an individual integration time of 10--50\,ms. 
A total of 100 cubes (i.e. 10$^5$ frames) and 50 cubes was acquired for HD\,130948 in the nights of July 25 and May 29, respectively, as well as flat and dark images for standard image correction. 
After finding the brightest speckle in each image of the cube, we can select an arbitrary percent of images based on the brightest speckles sorted {\it within the serie}. Hence, are selected those frames with the highest concentration of energy in a given speckle. 
A subsequent shift-and-add process of the previously selected frames products an image where the bright diffraction-limited core is surrounded by a fainter seeing halo. This technique is also referred as {\it Lucky Imaging} \citep{Law2006a}. The resulting angular resolution 
depends then on the percent of selected frames and, among others, on the natural seeing during the night and the integration time for each individual frame. 
The percent of selected frames results from a trade-off between a sufficiently high integration time and a good angular resolution. 
In our case, the individual integration time and the percent of selected frames were fixed, respectively, to 30\,ms and 30\%, 
leading to an effective total integration time of 900\,s. 
Thus, FastCam delivered $\sim$\,0.1$^{\prime\prime}$ resolution images, close to the diffraction-limit 1.22$\lambda$/D of 84\,mas of the NOT at 0.8\,$\mu$m. \\
As in any imaging system aiming at compensating the effect of the atmospheric turbulence, the presence of the seeing halo in the images is a strong limitation to high contrast capabilities. 
Since we are not able, with the current observational technique, to artificially reduce the starlight contribution as it can be done using a coronagraph, the flux contrast at a given distance from the star is improved by applying an additional post-processing stage to the image, which helps to suppress most of the seeing halo, unveiling fainter sources in the immediate vicinity of the star. 
We initially tried the classical solution of PSF-reference subtraction. However the strong time-variability of Lucky-Imaging PSF profiles at optical wavelengths prevents us from applying an efficient subtraction that would improve the detectability. Thus, we implemented a different approach based on the post-processing of the shift-and-add images in order to enhance particular spatial frequencies in the image. For this purpose, we have tested different image filtering algorithms, which are: \vspace{0.15cm} \\
$\star$ the subtraction of a median box filtering. \vspace{0.15cm} \\
$\star$ the implementation of a standard unsharp mask (i.e. $I$=$I$-$I$$\ast$$g$, where $I$ is the input image, ``$\ast$'' denotes convolution and $g$ is a Gaussian kernel. \vspace{0.15cm} \\
$\star$ the implementation of the wavelet transform of the shift-and-add image. All these algorithms, comparable to {\it edges enhancement} techniques, allow us to suppress a continuous offset (e.g. sky background, detector bias...) or an extended and diffused structure (e.g. a PSF halo) at a different spatial scale than the object of interest, hence favoring the detection of fainter point-like sources at higher spatial frequencies. \vspace{0.15cm} \\
The best results were obtained with the unsharp mask filtering and the wavelet decomposition. In their numerical implementation, those two algorithms are very similar, but the wavelet algorithm gives the advantage of a multi-resolution approach. 
This last solution was finally selected and applied to our two different dataset of HD\,130948, and is described in more details in Appendix~\ref{appendixA}.  

\subsection{Photometry}\label{primary-phot}

We used an absolute measurement of the HD\,130948\,A photometry to derive the $I$ magnitude of the companion from our images (see Sect.~\ref{companion-phot}). In order to determine the $I$ magnitude of HD\,130948\,A, we carried out a dedicated photometric measurement obtaining $I$-band images of HD\,130948 on 6 April 2010 using the CAMELOT instrument mounted on the IAC80  telescope. This optical camera consists of a 2k$\times$2k CCD detector with a 0.304$^{\prime\prime}$/pix plate scale providing a field of view of 10.4$\times$10.4 arcmin. The same $I$ filter can be physically interchanged in FastCam and CAMELOT and match the Johnson-Cousins system. The spectral responses of the each of the two detectors were obtained from available technical data and the differential effect on a solar-type star spectrum was estimated to a residual magnitude of 2.9\,milli-magnitude. We observed a serie of 10 images with individual exposure time of 5s. Raw data were reduced using routines within the IRAF environment. Bias-images were subtracted using the overscan region and zero exposure time images, and flat-field correction was applied using dome flats. We performed aperture photometry using routines from the DAOPHOT package. 
We adopted an aperture of 5 FWHM (depending on the seeing conditions, this number varies between 6.4$^{\prime\prime}$ to 8.7$^{\prime\prime}$) for our photometric standard stars from \cite{Landolt1992}. This is a standard aperture recommended by several photometric manuals to avoid contamination. Since the HD\,130948 images were defocused to avoid saturation, we adopted a larger aperture (10 FWHM) and hence, in order to correct for the difference, we estimated the aperture correction from 5 to 10 FWHM using bright and isolated Landolt standard stars. We obtained an aperture correction of 0.033 +/- 0.005 mag, which was included in the photometry of our target. 
Weather conditions during our observations were photometric as assessed by observing photometric standard all through the night, while average seeing ranged from 1.3 to 2$^{\prime\prime}$. 
In order to transform our instrumental magnitudes into apparent magnitudes, we observed four different Landolt standard star fields (each of them containing 3--6 standard stars) and repeated them along the night. We obtained 14 different images at 7 different pointings covering a range of airmasses from 1.1 to 2.1. We perform a linear fit to our data to obtain the zero points and the extinction coefficient, following the equation  $i-I$=$a_{\rm 0} + k\times airmass$, where $i$ and $I$ are, respectively, the instrumental magnitude and the apparent magnitude of the Landolt stars, $a_{\rm 0}$ is the zero point, $k$ is the extinction coefficient. We obtained $a_{\rm 0}$=2.68$\pm$0.019 (zmag=25) and $k$=0.17$\pm$0.013. The error bars in the calibration were obtained from the estimated errors of the coefficients in the linear fit. Eventually, our correction from instrumental to apparent magnitudes for HD\,130948 was finally $i-I$= 2.860$\pm$0.032 (-0.033$\pm$0.005 aperture correction ) = 2.827$\pm$0.032. The final error bar includes both the error in the calibration and in the instrumental magnitude. 

\section{Results}

\subsection{Detection of the BD companion}

\begin{figure*}
\centering
\includegraphics[width=6.0cm]{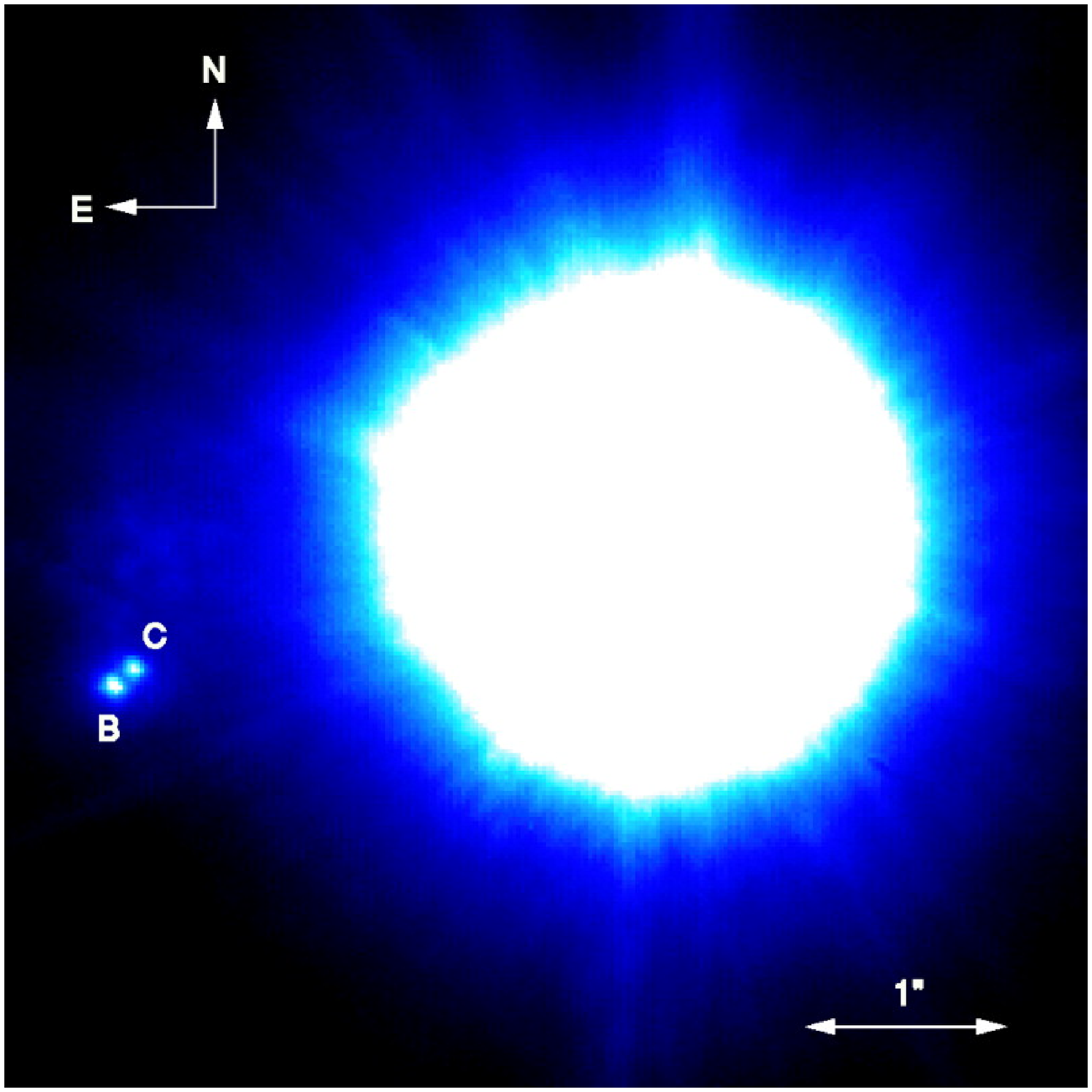}
\hspace{0.1cm}
\raisebox{0.0cm}{\includegraphics[width=6.0cm]{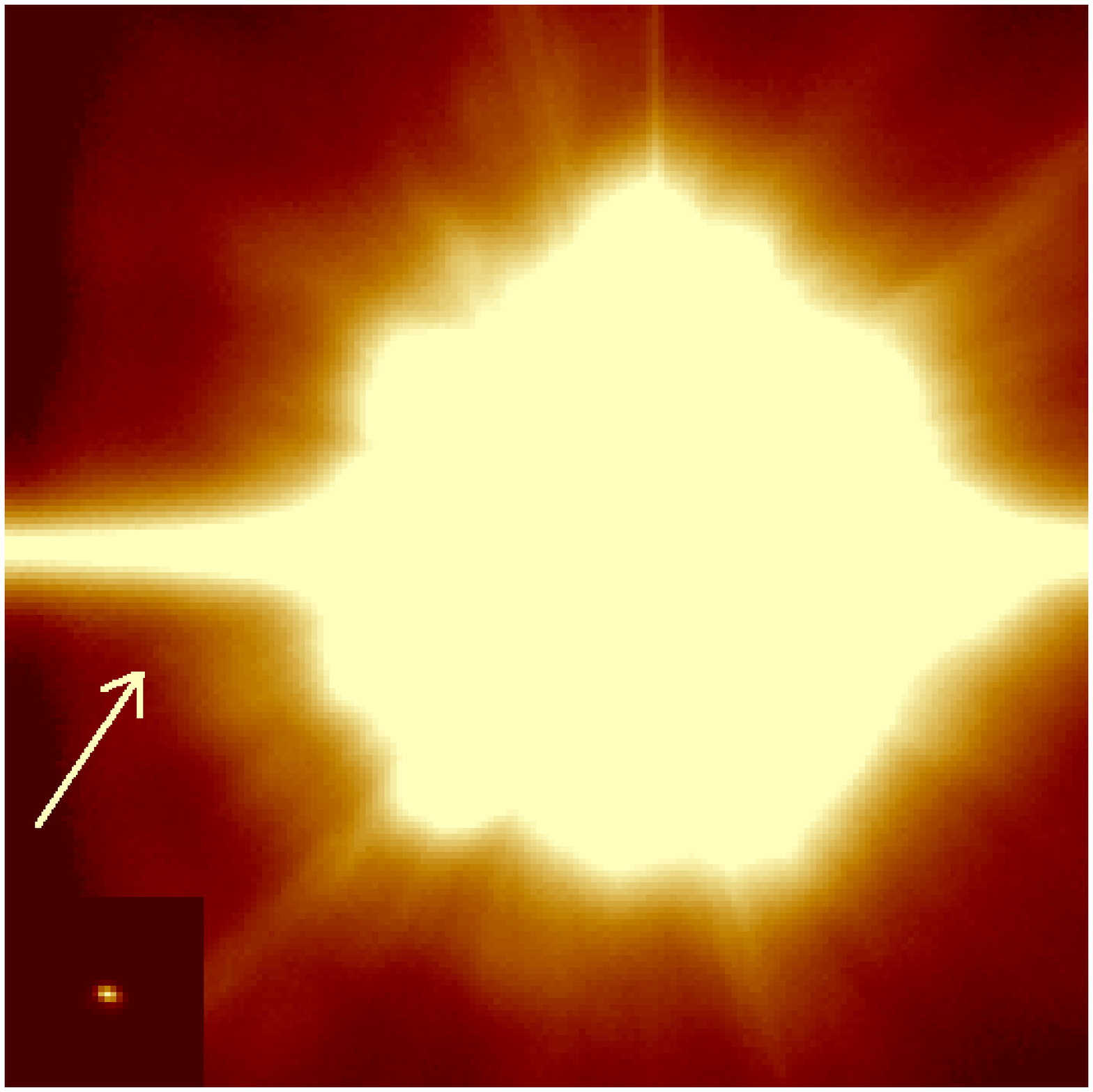}} \\
\vspace{0.1cm}
\raisebox{0.0cm}{\includegraphics[width=6.0cm]{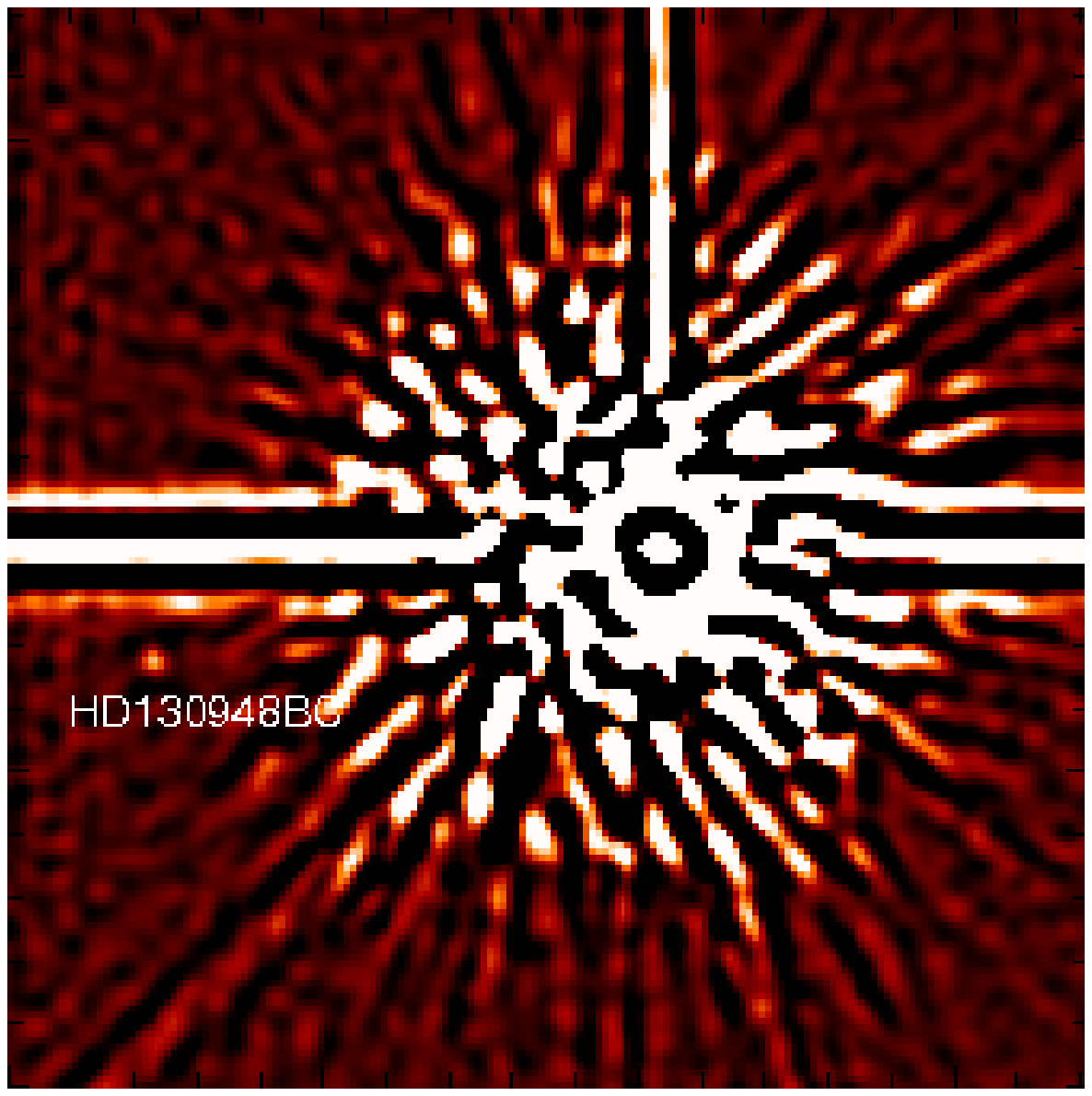}}
\hspace{0.1cm}
\raisebox{0.0cm}{\includegraphics[width=6.0cm]{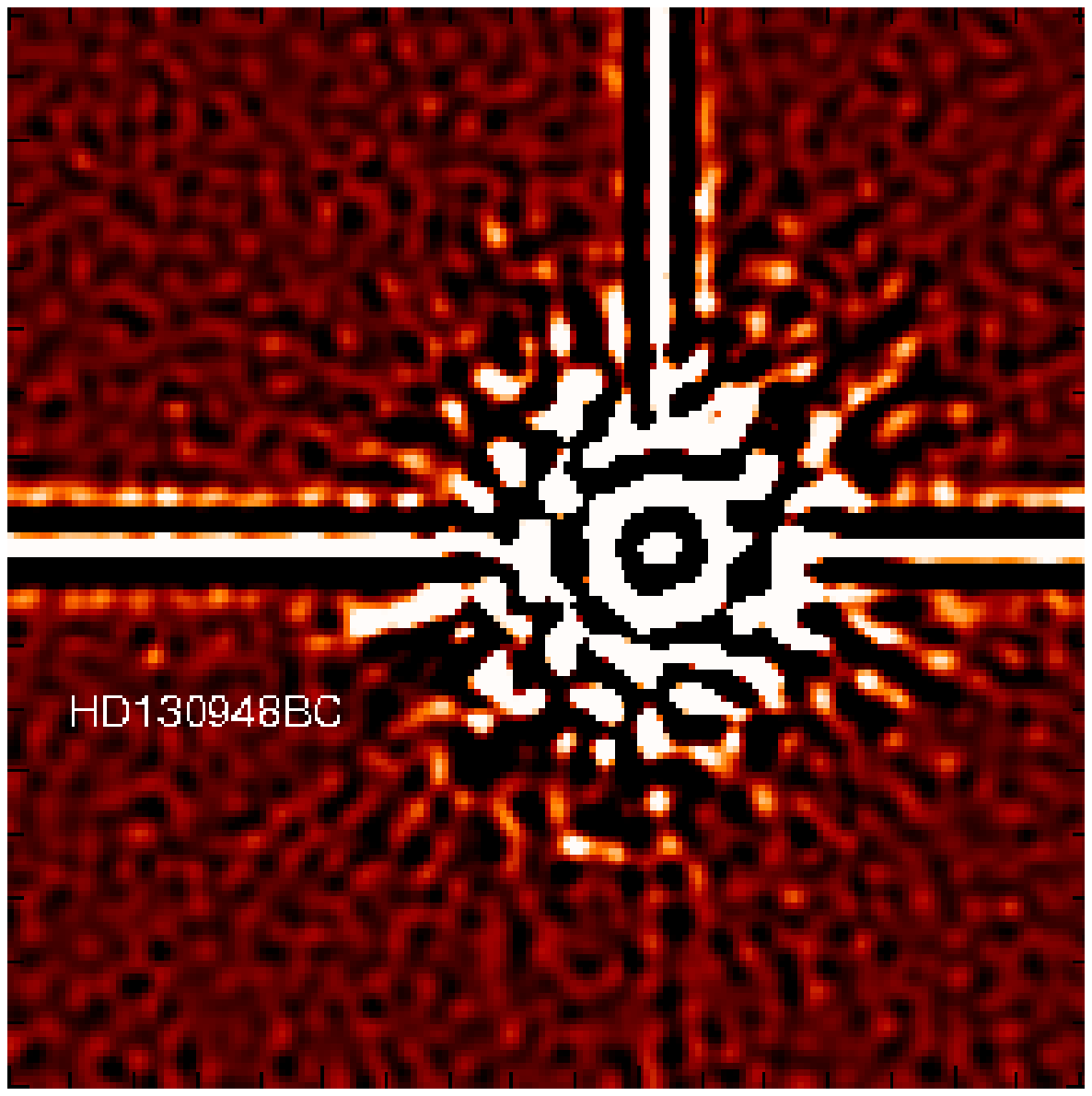}}
\caption{{\it Upper-Left:} Original $H$-band detection of HD\,130948\,BC in \cite{Potter2002}. -- {\it Upper-Right:} Direct lucky imaging image resulting from the best 30\% frames over a 10$^5$ serie. The white arrow indicates the anticipated position of the companion. The inset shows, on a different intensity scale, the core of the PSF. -- {\it Bottom-Left:} Post-processed image from July 2008 revealing the BD binary companion, unresolved with the NOT at 0.8\,$\mu$m. -- {\it Bottom-Right:} Post-processed image from May 2008 obtained with 5$\times$10$^4$ frames. All four images have the same size. North is up, East is left. The effective total integration time to obtain this image is 900\,s for the July image.}\label{image}
\end{figure*}

The panels of Fig.~\ref{image} show the imaging results from our observations. In the upper-left corner is displayed the original detection by \cite{Potter2002} in which the brown dwarf binary is resolved in the $H$-band with the 8-m Gemini-North telescope. The upper-right corner shows the direct shift-and-add image of HD\,130948 obtained with the data of July 25. 
In all the images, North is up, East is left. The FastCam images have been rotated by 90$^{\circ}$ with respect to the original position of the detector on sky. 
The average full-width-at-half-maximum (FWHM) is 131 mas, with a slight elongation in the East direction (148 mas) against the North direction (114 mas). 
We attribute this effect to atmospheric dispersion since no ADC (atmospheric dispersion compensator) is available in the current FastCam+NOT configuration. 
The white arrow indicates the expected position of the BD companion with respect to HD\,130948\,A, while the inset shows, on a different intensity scale, the core of the PSF. \\
\noindent The bottom part of Fig.~\ref{image} shows the HD\,130948 system observed in July (left) and May 2008 (right) after the image filtering step. The brown dwarf is detected with a signal-to-noise ratio (SNR) $\sim$9 and an average FWHM$\sim$110\,mas, 
at a position very consistent with earlier image (see next section). The BD binary \object{HD\,130948\,B} -- \object{HD\,130948\,C} is unresolved in our image. According to the orbital solution derived by \cite{Dupuy2009}, the BD separation at epoch 2008.56 is estimated to 30\,mas, below the theoretical diffraction-limit resolution of the NOT. The physical separation at the same epoch is 0.63\,AU assuming $d$=18.17\,pc given by the same authors.
\noindent Because the filtering process leaves speckles in the image that may mimic the presence of a companion, the detection is doubled-checked using a procedure that splits the 100 cubes of the July dataset into three separate datasets, which undergo the same post-processing step. In this way, speckle features simply resulting from the image processing are unlike to be repeated in the different sub-images, while a real companion remains detectable, although with a lower SNR. In our particular case, we benefit also from the additional dataset of May 29 (50 cubes) in which HD\,130948\,BC is searched for. In both cases, the BD companion binary is detected at the same location with respect to the G2V star. In the following part of this paper, the shift-and-add image of the best 30\% selected frames and the filtered image will be referred, respectively, as {\it Image 1} and {\it Image 2}.

\subsection{Detectability and contrast curve}

\begin{figure}[t!]
\hspace*{0.25cm}
\centering
\includegraphics[width=1.0\columnwidth]{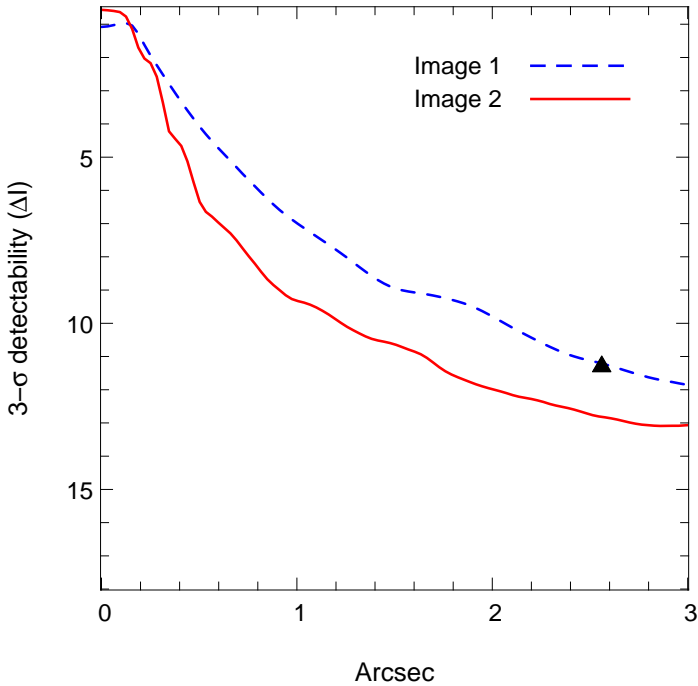}
\caption{
3-$\sigma$ detectability curve estimated in the two cases of Image\,1 (dashed line) and Image\,2 (continuous line). These data were obtained with 100 cubes and a 30\% best-frames selection per cube. The filled triangle shows the position of HD\,130948\,C (see text for details).}\label{contrast}
\end{figure}

In the 
Fig.~\ref{contrast}, 
we present the the 3-$\sigma$ {\it detectability} curve as a function of the distance to the central star for Image\,1 and Image\,2. These curves were obtained by estimating, in the corresponding images, the radial standard deviation profile in a 9$\times$9 pixels slipping box averaged in azimuth over 22 angular positions in clean areas of the image (i.e. not affected by the vertical and horizontal spikes visible in Fig.~\ref{image}) and separated by 5$^{\circ}$. The gain is of the order of 2 mag when implementing the post-processing step. 
This result, interpreted in the perspective of the images of Fig.~\ref{image}, suggests that the presence of the bright PSF halo in the direct image significantly contributes to the background noise at various spatial frequencies, degrading the detectability of a faint companion. On the contrary, in the filtered image the background noise associated to the bright halo is mostly filtered out around the spatial frequency constrained by the wavelet kernel. In other words, the wavelet filtering step strongly improves the signal-to-noise ratio of the object detection in absence of any external hardware system aiming at reducing the PSF halo (e.g. with a coronograph). 
In Fig.~\ref{contrast} was added a point showing the detection of HD\,130948\,BC (see next section for contrast estimation), which helps to better perceive the gain set by the post-processing step.

\subsection{Astrometry}

\noindent To determine the separation between the primary and the BD binary, we first used the photo-center of Image 1 (or barycenter in a 15$\times$15 pixels aperture centered on the primary) as the astrometric reference since we benefit from an excellent signal-to-noise in the unsaturated image of the primary. The position of the photo-center coincides with the brightest pixel within 0.09 pixels. The same result was obtained using a two-dimensional Gaussian fit of the primary PSF, which is explained by the high SNR obtained on this bright object. At the contrary, we only implemented the Gaussian fit technique to determine the position of the faint companion in the filtered image, which proved to be more accurate than the photo-center technique for lower SNR. 
We first checked, by simulating several artificial companions at similar distance and with comparable brightness to HD\,130948\,BC, that the presence of such a faint companion has negligible impact on the position of the primary photo-center. We also verified that the position of the faint companion {\it in the filtered image} (Image 2) is possibly altered by the process of spatial filtering by less than 0.2 pixels (or 0.007$^{\prime\prime}$) in average. The advantage of measuring the BD position in the filtered image is to benefit from an enhanced detection, removing the limitation of a poor signal-to-noise ratio. 
The two-dimensional Gaussian fit provided an estimation of the relative positions of the two components of 
2.561$^{\prime\prime}$$\pm$0.007$^{\prime\prime}$ at epoch 2008.56 (46.5\,AU at 18.17\,pc), slightly closer compared to the separation of 2.64$^{\prime\prime}$$\pm$0.01$^{\prime\prime}$ at epoch 2001.15 found by \cite{Potter2002}. The relative orientation of HD\,130948\,BC with respect to the primary is P.A.\,=\,102.7$^{\circ}$$\pm$0.15$^{\circ}$, to be compared with P.A.\,=\,104.5$^{\circ}$$\pm$0.5$^{\circ}$ found by the previous authors. Assuming the distance revised by \cite{Dupuy2009} to $d$=18.17$\pm$0.11\,pc, these measurements may indicate an orbital motion around the primary, with a projected physical separation decreased by 1.43$\pm$0.22\,AU in 7.4\,yr and a P.A. change by 1.8$^{\circ}$$\pm$0.55$^{\circ}$ in the same period. These values are also in agreement with the astrometric data provided by \citet{Geissler2008} in their mid-infrared observation of the HD\,130948 system at epoch 2006.51. This change corresponds to an approximate orbital velocity projected onto the sky of $\sim$1\,km.s$^{-1}$, to be compared with an unprojected orbital velocity of $\sim$4.6\,km.s$^{-1}$ assuming a circular orbit of radius 46.5\,AU and a stellar mass of 1.11\,$M_{\odot}$ for HD\,130948\,A. The position of the object has been also measured in the image from May observations. Although the estimation is less precise ($\pm$0.015$^{\prime\prime}$ and $\pm$0.3$^{\circ}$), we find consistent values between the two epochs with 2.56$^{\prime\prime}$ and 102.5$^{\circ}$ at epoch 2008.41.

\subsection{Photometry of HD\,130948\,BC}\label{companion-phot}

\begin{figure}[b]
\hspace*{0.25cm}
\centering
\includegraphics[width=0.8\columnwidth]{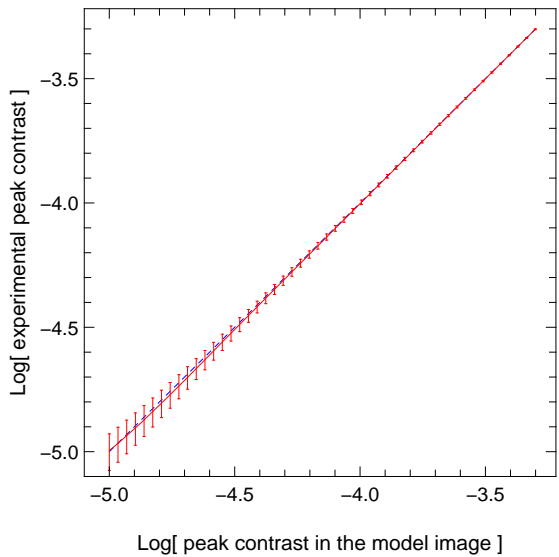}
\caption{Evaluation of the impact of the image post-processing on the peak value of a nearby companion. The solid curve with error bars shows, in the y-axis, the experimental mean and rms values of the measured peak contrast after image post-processed for various simulated companions contrast at a distance comparable to HD\,130948\,BC as an input. The x-axis shows the input pre-determined model contrast. The dash line is the $y$=$x$ curve and represents for comparison the contrast values ideally unaffected by the post-processing stage.}\label{contraste3}
\end{figure}

The photometry of the BD companion is extrapolated as a relative measurement to the primary, which is described in Sect.~\ref{primary-phot}. 
From the data obtained with the CAMELOT instrument, we have measured the $I$-band magnitude of HD\,130948\,A to be 5.19$\pm$0.03. \vspace{0.15cm} \\
The estimation of the contrast between HD\,130948 and the BD companion was obtain by relative comparison with a set of 25 fake companions simulated with comparable separation and brightness in an image model. We compared both the peak values of the companions and the fluxes summed in a small aperture of 3-pixels in radius.\\
\noindent In order to evaluate the effect of the image post-processing on the BD peak value, 
the contrast of the fake companions in the image model is spanned into 50 values ranging from 5$\times$10$^{-4}$ to 10$^{-5}$. 
Each of these 50 models is convolved with our observed PSF of HD\,130948 (Image\,1) and then filtered, which permits us to directly compare, in the processed image, the peak values between the synthesized companion and the original detection. The significance of this approach is justified by the small width of the field-of-view ($\pm$3$^{\prime\prime}$), which is within the typical isoplanatic patch. 
As a consequence, the speckle pattern in the raw short-exposure is not expected to show significant spatial variations within $\pm$3$^{\prime\prime}$, with comparable implications on the PSF of Image\,1. 
Hence, for each contrast value in the image model, we benefit from 25 independent experimental contrasts from which are extracted mean and rms values. The statistical dispersion express possible PSF distortion effects in the field-of-view affecting the peak value of the companions (see Fig.~\ref{contraste3}). From this empirical relationship, the contrast of HD\,130948\,BC is estimated by interpolation to the measured peak value. As a second approach we have also applied the reverse reasoning, which consists in tuning to an arbitrary precision the fake companion contrasts {\it in the image model} to match the experimental peak value of HD\,13948\,BC in the post-processed image. We then derive a new estimate of the mean contrast and corresponding dispersion. 
Eventually, we apply the same procedures using as input measurements the fluxes summed in the aperture described above. \\ 
We are provided in the end with four different estimates of HD\,130948\,BC contrast with a conservative value of 3.0$\pm$0.3$\times$10$^{-5}$ ($\Delta I$=11.30$\pm$0.11) with respect to the $I$=5.19 primary. Note that the peak value method provides a better accuracy ($\pm$0.21$\times$10$^{-5}$) than the aperture photometry ($\pm$0.32$\times$10$^{-5}$), in part because in the last case we do not weight the contribution of each aperture pixel as this can be done with optimal extraction techniques. 
%
%
%
%
The conversion into magnitudes results into an unresolved BD magnitude $I$=16.49$\pm$0.11 (or $I$=17.24 per component assuming equal brightness for the two components). The combined MKO photometry of the BD from \cite{Dupuy2009} results in the $J$=13.2$\pm$0.08 magnitude, leading to a colour index $I$-$J$=3.29$\pm$0.13 for the unresolved binary. We have also verified that the possible elongation by $\sim$30\,mas of the unresolved binary has a negligible effect ($\lesssim$5\,milli-mag) on the measured photometry.

\section{Discussion}

\subsection{Prospects for the technique} 

\begin{figure}[t]
\centering
\includegraphics[width=8.0cm]{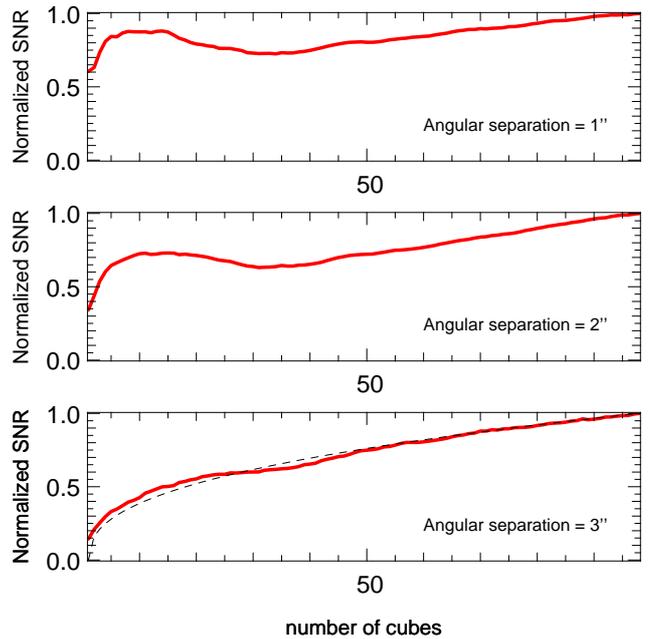}
\caption{
Normalized 
SNR 
as a function of the number of co-added cubes used for the final post-processed image. The curves were obtained by computing the spatial rms in a 20$\times$20 pixels box at 3$^{\prime\prime}$, 2$^{\prime\prime}$ and 1$^{\prime\prime}$ from the primary. As a reminder, in the present data one cube is composed of 1000 images with 30\,ms exposure from which are selected the 30\% ``best'' (brightest speckle) images. We overplotted the power-law that best reproduce the experimental data for 3$^{\prime\prime}$ separation, namely a N$^{-0.4}$ law (dotted line). The x-axis starts at value 0, and each sub-interval represents 5 cubes.
}\label{snr-curve}
\end{figure}

In our present work, we apply the wavelet filtering step to images that do not benefit from AO-assisted or coronographic techniques. 
We anticipate that the assistance of, at least, a moderate AO correction able to reduce the halo contribution would improve the achievable contrast at shorter radii in comparison to what we have obtained so far. Hence, we plan to further explore the potential of our approach with observations using the 4.2-m WHT and its AO-system NAOMI. 
A fundamental aspect we investigated here to illustrate the potential of the technique is the expected sensitivity limit as a function of the distance to the central star. 
We have selected reasonably clean regions (i.e. unaffected by optical artefacts) of 20$\times$20 pixels distributed over more than 10 azimuthal positions, centered at 3$^{\prime\prime}$, 2$^{\prime\prime}$, 1$^{\prime\prime}$, respectively, from the center of Image 2. We have then estimated an average of the spatial standard deviation as a function of the number of co-added cubes. The result is illustrated in Fig.~\ref{snr-curve} with the evolution of normalized SNR 
as a function of the number of co-added cubes. The bottom plot shows a regular increase of the normalized SNR 
following a $N^{-0.4}$ power law, where $N$ is the number of co-added cubes. 
Closer to the star, the plots at 2$^{\prime\prime}$ and 1$^{\prime\prime}$ obtained with the current data indicate that a relatively high SNR is quickly obtained already after $\sim$10 cubes, while only slowly increasing afterwards compared to the 3$^{\prime\prime}$. Although it is difficult to derive a clear and decisive trend simply with the current data, these plots together with the filtered images presented in this work suggest that as close as 1$^{\prime\prime}$--2$^{\prime\prime}$ the sensitivity is limited by residual speckles, which dominate the image as for traditional AO. Hence, increasing the number of cubes under this regime does not provide a significant gain. This effect should become limited to smaller separation with telescopes larger than the 2.5-m NOT. The small drop in the top plot of Fig.~\ref{snr-curve} around $N$=35 is possibly due to an unexpected increase of the speckle noise contribution in the corresponding images. 
Being increasingly dominated by speckle noise at short distance from the star, a further study of the speckle pattern statistics would be required, which is nevertheless not within the objectives of the current paper.

\subsection{Colour}

\begin{figure}[t]
\centering
\includegraphics[width=8.5cm]{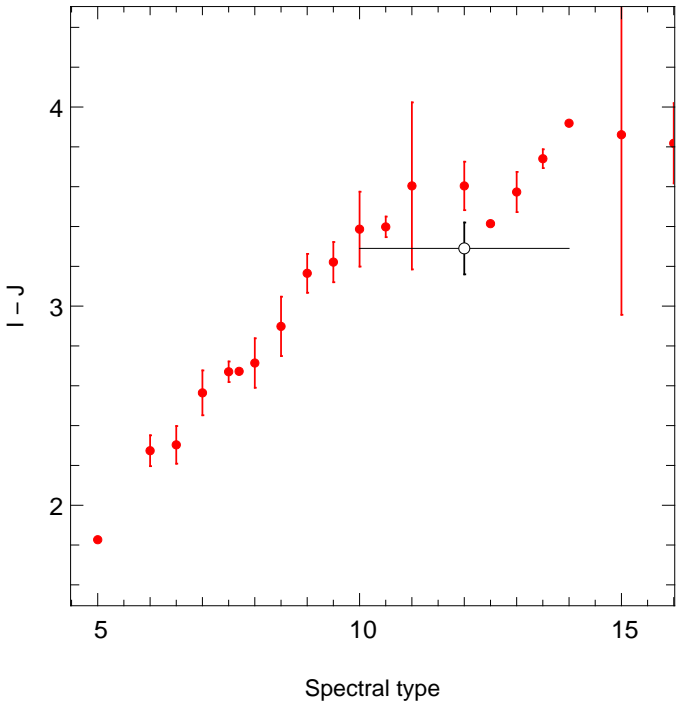}
\caption{
$I_c$-$J_{\rm MKO}$ colour versus spectral type reproduced from \citet{Liebert2006} for late M and L dwarfs. 
Spectral types 5 to 15 corresponds to M5 to L5. 
Error bars in the photometry represent the measurement dispersion (rms). 
Our measurement at the 1-$\sigma$ confidence level of HD\,130948\,BC colour is shown as open circle. 
}\label{color-plot}
\end{figure}

We place our colour measurement in the context of L-type brown dwarfs available photometry \citep{Liebert2006}. In Fig.~\ref{color-plot}, we reproduce a plot from these last authors 
giving the $I_c$-$J_{\rm MKO}$ colour versus spectral type for a sample of late M and L dwarfs, after conversion from the 2MASS to the MKO photometric system. 
Our first measurement of the integrated $I_c$ magnitude of the HD\,130948 BD companion allows us to compare it with existing data. We assume for HD\,130948\,BC a spectral type L2$\pm$2 inferred by near-infrared spectroscopy by \citet{Potter2002}. 
The $I$-$J$ colour mean and dispersion in the estimated spectral type interval for the \cite{Liebert2006} sample equal 3.58$\pm$0.17. 
Within the $\pm$2 sub-class uncertainty, the given sample shows only one L dwarfs over 21 objects with bluer colour ($I$-$J$=3.19) than HD\,130948\,BC, and classified L0. Hence, this suggests that our object clearly lies in the blue region of the colour distribution for this spectral range.  
This trend becomes even stronger if we place our measurement in the context of the work by \cite{Goto2002}, who classified HD\,130948\,BC as L4$\pm$1. 
A similar behavior is also found when placing our colour measurement in the $I$-$J$ versus spectral type plots in \cite{Dahn2002} and \cite{Bihain2010} 
, where typical colours $\sim$\,3.4--3.6 are found for L2 dwarfs. 
The field dwarf \object{2MASSW\,J1841086+311727} (L4-type, $I_c$-$J$$\approx$3.2, not plotted in Fig.~\ref{color-plot}) is another known case of spectroscopically reported as an L4 dwarf, although it appears too blue even for an early L \citep{Kirkpatrick2000}. 
Eventually, our BD object was placed in the $J$-$K$ versus $I$-$J$ plot of \cite{Liebert2006} using $J$-$K$$\sim$1.6 measured by \cite{Bihain2010}. In this plot, HD\,130948\,BC appears consistent with an L2 spectral type, although lying slightly off the bulk of the L2 BD distribution, suggesting that the object may be peculiar in that diagram as well. 
Considering this trend, we intend to provide some qualitative explanations to the relative blue colour of our object, keeping in mind that further observations would certainly help investigating them. \\
As suggested by \cite{Marley2002}, a possible reason is that for late L dwarfs, the shorter wavelength bands are influenced more by the wings of potassium and other alkali gas opacities. \\
Also, the metallicity of HD\,130948 was measured by \cite{Valenti2005} to Fe/H=0.050$\pm$0.030, i.e. solar metallicity, thus presumably the same value can apply to the BD companion. L-type objects in the Pleiades, with metallicity comparable to the Sun, do not present any remarkable bluer $I$-$J$ colours \citep{Bihain2010}, hence metallicity can probably be ruled out as a cause of the observed effect. \\
Low surface gravity could also result in a bluer $I$-$J$ colour due to a weaker potassium line \citep{Kirkpatrick2008}. Since our measured $I$-$J$ colour for HD\,130948 is slightly smaller than the average $I$-$J$ colour in the $\sim$120\,Myr Pleiades (cf. Fig.~5 in \cite{Bihain2010}), we may have an indication for an even younger age of HD\,130948 in comparison to the 0.1--1\,Gyr range of age proposed by \cite{Dupuy2009}. 
Also, weather-like phenomena in the BD atmosphere \citep{Bailer1999,Goldman2008} could cause flux time-variability resulting in a larger dispersion of the $I$-$J$ colour. 
Such a case may be explored with repeated monitoring observations of the object, which could be possibly conducted using the observational technique investigated in this paper. Alternatively, stable differences in the surface cloud coverage or in the atmosphere dust properties could alter 
the measured $I$-$J$ index.\\
An additional hypothesis, related to the presence of dust in the neighborhood of HD\,130948\,BC system, could be formulated as the effect of stellar radiation scattered by a residual of dusty disk around the BD companion, with an increased efficiency of the phenomenon towards optical wavelengths \citep{Kalas2007}. Using modeling tools such as the {\it debris disk radiative transfer simulator}\,\footnote{www1.astrophysik.uni-kiel.de/dds/} (DDS, \citet{Wolf2005}), we were able to assess that a small-size ``circum-brown dwarf'' disk, with a $\sim$10$^{-7}$$M_{\oplus}$ mass and composed of small ($\sim$0.2\,$\mu$m) grains confined in an $\sim$2\,AU region in radius around the BC companion, would produce a $I$-$J$ variation of $\sim$0.2 mag towards bluer colours. However, the effective extent and geometry of such a disk can clearly not be constrained with these current data. Further analytical and observational tests (e.g. focusing on searches for small mid-IR excess for this object) should indicate us if such a compact dusty disk would survive in a relatively old multiple system, and how the binary nature of HD\,130948\,BC, with a $\sim$2.2\,AU semi-major axis \citep{Dupuy2009}, or even the interaction with the primary would dynamical sculpt the dust distribution into a circum-(substellar)binary \citep{Artymowicz1994} and/or circum-(sub)stellar \citep{Mathieu2007} component. Note that, whether this hypothesis is plausible or not, the presence of a disk would not be uncommon in substellar objects \citep{Zapatero2007,Luhman2009}. 

\section{Conclusions}

This work addresses the question of high contrast in optical speckle imaging in the context of substellar objects, in a spectral domain where conventional AO systems 
present modest performances for long-exposure imaging. 
We show that the Lucky-imaging approach has a significant potential for the detectability of faint and close companions to bright stars, which can be improved with an additional 
stage of post-processing based on the image wavelet transform. We successfully applied this approach for the first time to the system HD\,130948 with a clear detection in the $I_c$ filter of the BD binary companion HD\,130948\,BC at $\sim$\,2.5$^{\prime\prime}$ using the 2.5-m NOT telescope, and in the absence of any coronographic system. The BD companion is spatially unresolved because of a too small physical separation (30\,mas) at the time of our observations, which is below the telescope resolution at this wavelength. The relative contrast at this distance is estimated to $\Delta I$\,=\,11.30 ($\sim$3.0$\times$10$^{-5}$). At 1$^{\prime\prime}$, the point-source detectability is estimated at $\Delta$$I$$\sim$10 in the neighborhood of a bright star like HD\,130948\,A, making the proposed approach a powerful technique to obtain high resolution and high-contrast photometry even with small class telescopes. 
Considering these results, we plan hopefully a positive extension of this approach to larger and AO-assisted medium class telescopes \citep{Law2009} or eventually to coronographic imaging system. Alternatives to substellar programs could also be considered for science cases where optical imaging is an asset, as for instance in the case of the search for white dwarf companions to bright stars.

\appendix
\section{Comparison and details on different image processing algorithms}\label{appendixA}

In this work, three different algorithms to enhance the detection of HD\,130948\,BC. For all of the three options, the filters are applied recursively in order to improve the companion detection ($\sim$3 iterations, see below). These algorithms are: \vspace{0.15cm} \\
$\star$ a median box filter, $I$=$I$-median($I$,N), where N is the size of the median box filter, followed by a convolution with Gaussian kernel to filter out pixel noise. \vspace{0.15cm} \\
$\star$ an unsharp mask filter, $I$=$I$-$I$$\ast$$g$, where $g$ is a Gaussian kernel, followed by pixel noise filtering. \vspace{0.15cm} \\
$\star$ a wavelet mask filter resulting from the wavelet transform of the image. As a multi-resolution decomposition of the original image, this transform provides a set of filtered images from which can be selected the best trade-off between the spatial resolution and the SNR. The basic algorithm used in this work is largely inspired from the wavelet ``\`a trous'' procedure written in C/Yorick language and part of the Yeti package (see {www-obs.univ-lyon1.fr/labo/perso/eric.thiebaut/yeti.html}). We apply to the dataset $a$ the function WT($a$,ORDER), where WT is the wavelet transform described hereafter in the one-dimensional case and extendable to the two-dimensional case of images for our work. The parameter ORDER gives the maximum degree of decomposition in the multi-resolution analysis. 
\begin{figure}[b]
\centering
\includegraphics[width=2.8cm]{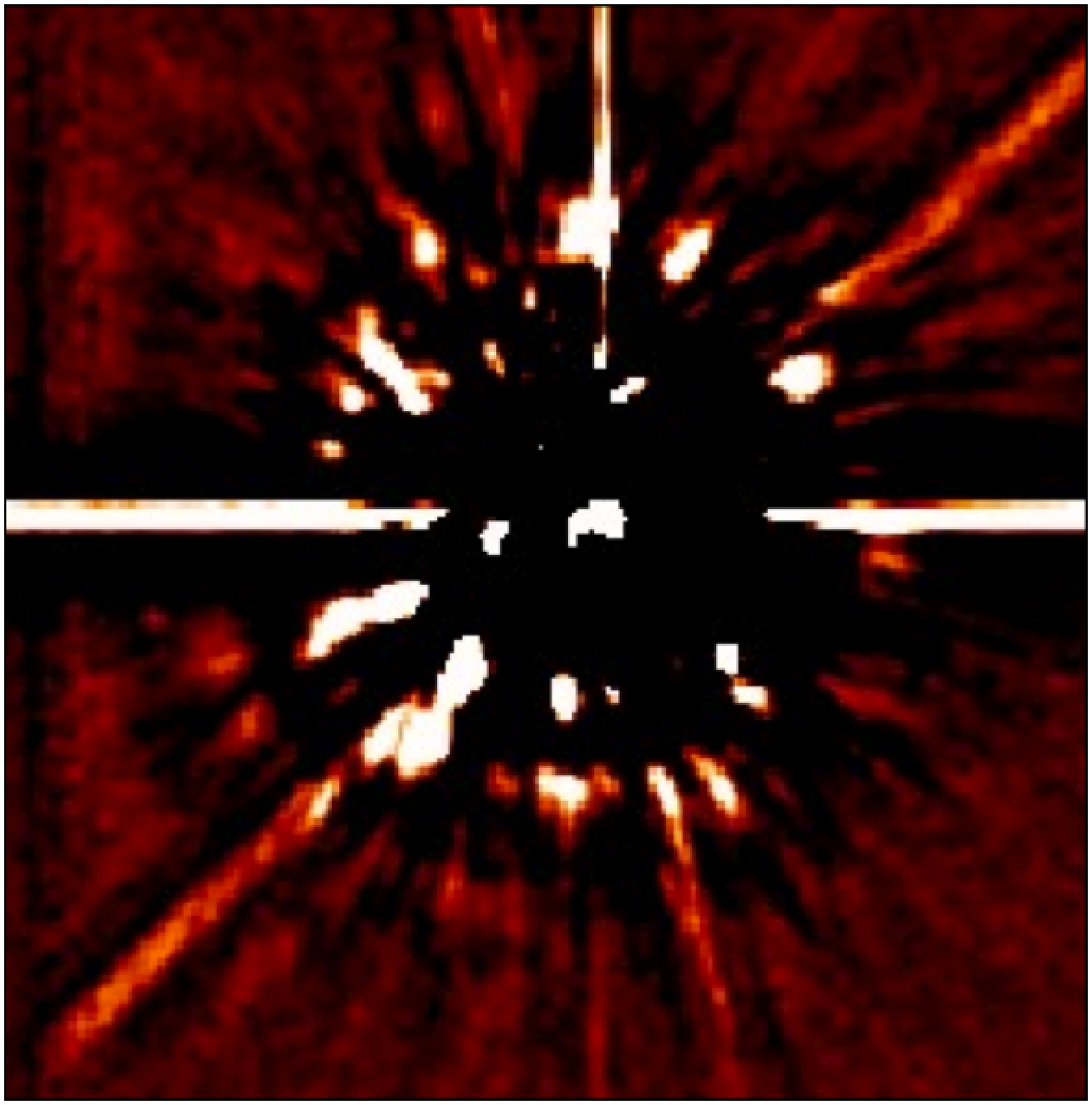}
\hspace{0.05cm}
\includegraphics[width=2.8cm]{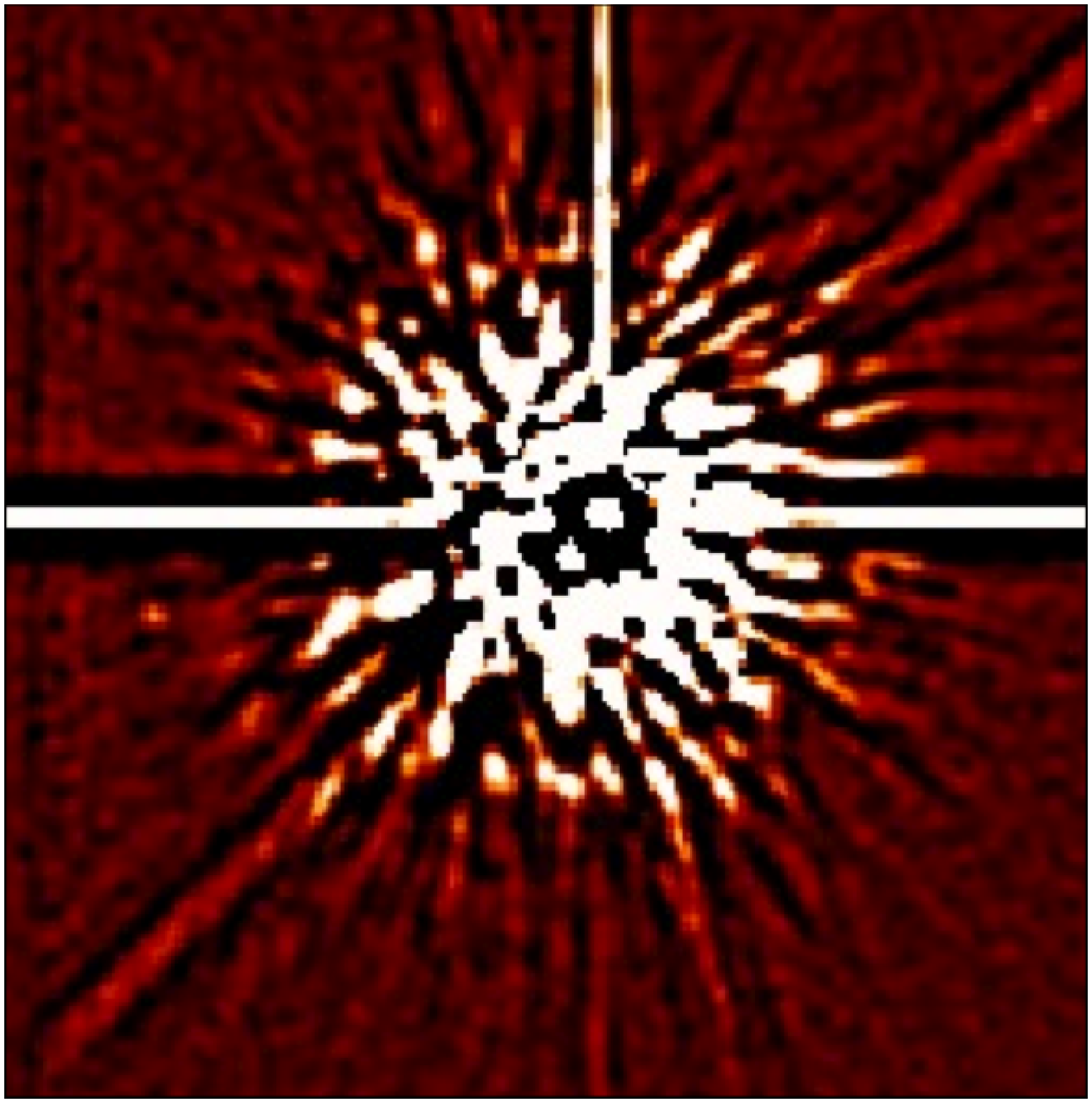}
\hspace{0.05cm}
\includegraphics[width=2.8cm]{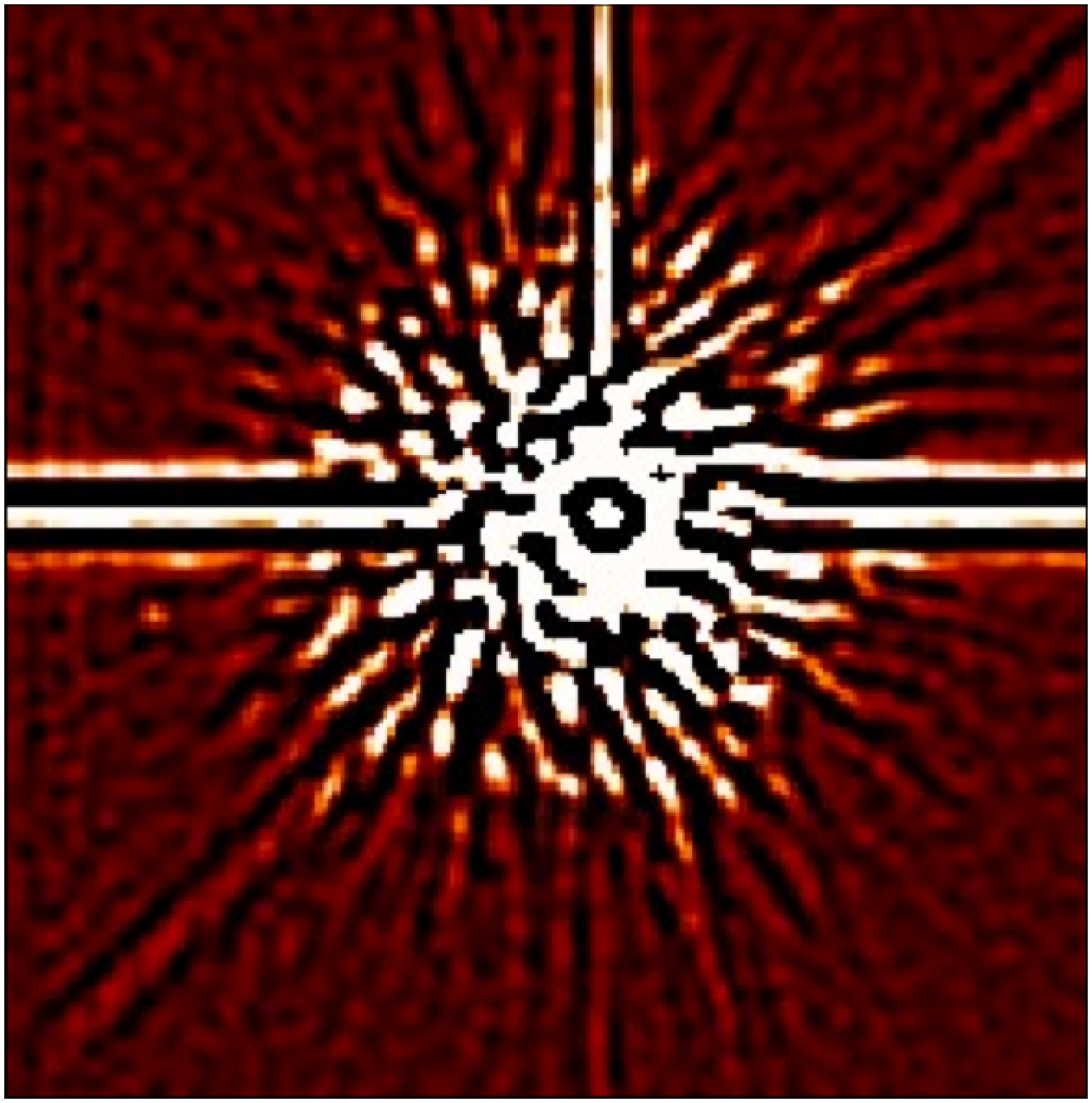}
\caption{Effect of the recursivity in the image filtering process. This serie shows our resulting filtered image of the HD\,130948 system after one (left), two (center) and three (right) iterations of the wavelet algorithm used in this case.}\label{myiteration}
\end{figure}
In the following, $i$ is the intermediate decomposition degree (also called order) and 1$\leqslant$$i$$\leqslant$ORDER+1. 
By applying the operation $b$=WT($a$,ORDER), $b$ becomes a cube of images of the same size as $a$ and with its third dimension equal to ORDER+1, as explained hereafter. The image $b_{\rm i}$ of intermediate order $i$ is given by:
\begin{eqnarray}
&& a_{\rm 1}  =  a \\
&& b_{\rm i}  =  a_{\rm i} - a_{\rm i+1} \\
&& a_{\rm i+1}  =  T(a_{\rm i}, {\rm SCALE}=2^{\rm i -1})
\end{eqnarray}
$T$ is a convolution operation by:  
\begin{eqnarray}
&& a_{\rm i+1}(p) = \sum_{-W \leqslant j \leqslant W}k(j+W+1)\times a_{\rm i}(p+j\times{\rm SCALE})
\end{eqnarray}
where SCALE=2$^{i-1}$. Integer $p$ is the pixel number in the image, ``SCALE'' and $W$ are, respectively, the scaling factor and the half-size of the wavelet kernel $k$ (i.e. $k$ has a length of 2$\times$$W$+1 pixels in the one-dimensional case). From cube $b$, we selected one filtered image, at the optimal order which represents the best trade-off between the necessary spatial resolution to detect the companion and a good SNR. If required, the selected filtered image becomes the input image $a$ and the filtering process is run recursively, keeping the same value of the ORDER parameter and eventually selecting the same optimal intermediate order. 
The reason why this process is run recursively -- or iteratively -- is to achieve a higher level of suppression of remaining ``low spatial frequency structures'' in the image, as illustrated in the sequence of Fig.~\ref{myiteration}. We verified empirically that there is generally no significant gain after a maximum of 3 iterations.
\begin{figure}[t]
\centering
\includegraphics[width=8.8cm]{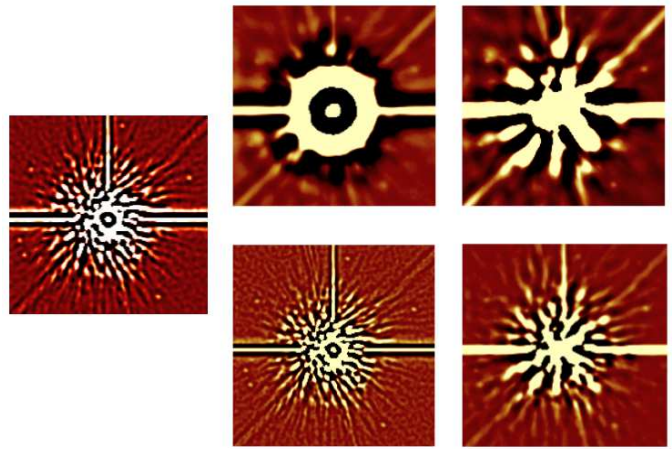}
\caption{Comparison of the different algorithms used in this study. The intensity scale has been optimized to detect the fake companions. 10 individual companions have been simulated for the tests. Only 5 are displayed here. All the presented images have undergone 3 iterations of the tested filtering algorithms. {\it Left:} wavelet algorithm used in this work. 
{\it Upper-middle:} 20$\times$20 pixels Gaussian kernel filtering followed by 3$\times$3 Gaussian smoothing to remove pixel noise. 
{\it Lower-middle:} 5$\times$5 pixels Gaussian kernel filtering followed by 3$\times$3 Gaussian smoothing to remove pixel noise. 
{\it Upper-right:} 20$\times$20 pixels median filtering followed by 3$\times$3 Gaussian smoothing to remove pixel noise. 
{\it Lower-right:} 5$\times$5 pixels median filtering followed by 3$\times$3 Gaussian smoothing to remove pixel noise.}\label{comparaison-algo}
\end{figure}\\
\noindent The wavelet kernel is support-limited and has been empirically optimized for this work to a 5$\times$5 pixels square (i.e. $W$=2) with a two-dimensional distribution given by 
[[1,1,1,1,1],[1,3,3,3,1],[1,3,8,3,1],[1,3,3,3,1],[1,1,1,1,1]]\vspace{0.15cm}.\\
In order to evaluate the effects of each algorithm, we have created 10 fake companions with similar properties to HD\,130948\,BC and radially distributed around HD\,130948\,A. As an additional test, this permitted us to discard possible algorithm artefacts that would have led to a false detection of HD\,130948\,BC.\\
The imaging results are shown in Fig.~\ref{comparaison-algo}. We initially implemented the two first algorithms using a kernel with a FWHM of 4--5 times the original PSF FWHM. In both cases, this has led to an excessive smoothing of the image preventing us from detecting all of the companions and in particular HD\,130948\,BC (see Fig.~\ref{comparaison-algo}, upper-middle and upper-right panel). 
Reducing the convolution kernel to a size comparable with the PSF FWHM improved significantly the result, although the median box filter does not reach a satisfactory level. On the contrary, the iterative unsharp mask filtering using a smaller kernel permits us to detect all the companions, included HD\,130948\,BC. This suggests that, apart from the multi-resolution decomposition -- which is useful to identify a close-to optimal kernel size -- unsharp mask filtering is almost identical in terms of point-like source detectability to the wavelet filtering algorithm described above.

\begin{acknowledgements}
LL is funded by the Spanish MICINN under the Consolider-Ingenio 2010 Program grant CSD2006-00070: First Science with the GTC (www.iac.es/consolider-ingenio-gtc). 
We wish to thank our anonymous referee who helped us to significantly improve the manuscript. 
Authors thanks 
support astronomer R. Barrena and the IAC maintenance team. Based on observations made with the Nordic Optical Telescope 
in the Spanish Observatorio del Roque de los Muchachos of the IAC. This work made use of Yorick (www.maumae.net/yorick), 
and of the SIMBAD database operated at CDS, Strasbourg, France. 
\end{acknowledgements}

\bibliographystyle{aa}
\bibliography{paper_gc.bib}

\end{document}